\documentclass[3p]{elsarticle}

\usepackage{lineno,hyperref}
\modulolinenumbers[5]

\journal{arXiv.org}

\renewcommand{\vec}[1]{\ensuremath{\mathbf{#1}}}
\begin{document}

\begin{frontmatter}

\title{Controlling bubble coalescence in metallic foams: A simple phase field-based approach}

\author{Samad Vakili \fnref{fn1}}
\ead{samad.vakili@rub.de}

\author{Ingo Steinbach}
\ead{ingo.steinbach@rub.de}

\author{Fathollah Varnik\corref{cor1}}
\ead{fathollah.varnik@rub.de}

\cortext[cor1]{Corresponding author}

\address{Ruhr-Universit\"at Bochum, Interdisciplinary Center for Advanced Materials
Simulation (ICAMS), Universit\"atsstr. 150, 44801 Bochum, Germany}

\fntext[myfootnote]{First author}

\begin{abstract}
The phase-field method is used as a basis to develop a strictly mass conserving, yet simple, model for simulation of two-phase flow. The model is aimed to be applied for the study of structure evolution in metallic foams. In this regard, the critical issue is to control the rate of bubble coalescence compared to concurrent processes such as their rearrangement due to fluid motion. In the present model, this is achieved by tuning the interface energy as a free parameter. The model is validated by a number of benchmark tests. First, stability of a  two dimensional bubble is investigated by the Young-Laplace law for different values of the interface energy. Then, the coalescence of two bubbles is simulated until the system reaches equilibrium with a circular shape. To address the major capability of the present model for the formation of foam structure, the bubble coalescence is simulated for various values of interface energy in order to slow down the merging process. These simulations are repeated in the presence of a rotational flow to highlight the fact that the model allows to suppress the coalescence process compared to the motion of bubbles relative to each other. 
\end{abstract}


\end{frontmatter}

\section{Introduction}
\label{sec:Int}

The aim of this work is to develop a new model for simulation of two-phase flow to be applied in the microstructure evolution of metallic foams. In these materials, before the solidification, a large number of bubbles is densely distributed inside the melt. Therefore, in the processing stage where there is a flow inside the system, the contact of the bubbles and their coalescence are inevitable \cite{Banhart2006}. However, coalescence rate must be kept low compared to the other simultaneous processes. Otherwise, the stability of the foam is broken and no foam structure would be expected \cite{Langevin2000}. Reviewing the literature, there are different models for simulation of the foam structures. Some are based on the simplification to neglect the dynamics of gas phase and only consider its pressure by imposing the pressure boundary conditions \cite{Koerner2002,Bueno2016, Anderl2014, Yin2006}. Other models consider both gas and liquid dynamics and nevertheless impose simplifications to create the foam structure \cite{Uehara2014}. Furthermore, in order to stabilize the foam structure, surfactants are introduced to create the interface energy gradient in order to control the coalescence of bubbles as a result of Marangoni's effect \cite{Bueno2016, Anderl2014}. Here, we propose a model which resolves the fluid dynamics inside all bulk phases. At the same time, we do not explicitly introduce surfactants but control the rate of coalescence by treating the interface free energy as a free tunable parameter. This is performed without changing the bulk properties of the gas and the liquid phases. As will be shown below, the model has the capability to suppress the rate of bubble-coalescence compared to other concurrent processes such as rearrangement of bubbles due to rotational motion. Therefore, it can be utilized to study the time evolution of foam structures.\\
\\
The developed model is based on the phase-field method, which is widely used in the modeling of different physical phenomena involving the interface dynamics, such as dendritic growth \cite{Steinbach1999b}, Hele-Shaw flows \cite{Caginalp1989}, and two-phase flows \cite{Jacqmin1999}. The basis of this approach is on diffuse interface models, where the interface is defined to have a finite width. In phase-field methods for two-phase flow \cite{Jacqmin1999, Anderson1998, Jamet2001}, the interface dynamics is obtained by using a conserved quantity, as an order parameter, which involves a fourth order differential equation. For the case of phase-field method using a non-conserved order parameter \cite{Steinbach1999b}, the interface dynamic equation reduces to second order, which is computationally favorable.\\
\\
In the current model, the phase-field $\phi$ is coupled with the density $\rho$ to address the dynamics of two immiscible fluids. However, unlike the concentration field in the solidification of alloys \cite{Steinbach2013}, the density in the present work is not an independent parameter and does only vary as a function of the volume occupied by a phase (integral of $\phi$). The work is organized as follows. In Sec.~(\ref{sec:PressureTensor}), an interfacial force is obtained for a general form of free energy functional. Then, in Sec.~(\ref{sec:Squ}), the specific form of this force is derived for a square gradient model. The connection between the unknowns of the square gradient model and the physical parameters of the system is provided in Secs.(\ref{subsec:Sta_bubble}) and~(\ref{subsec:Int_h}). Sec.~(\ref{sec:Model}) describes fluid dynamical equations and their coupling to the phase-field dynamics. The model is then applied in Sec.~(\ref{sec:Res}) to a number of benchmark simulations. Most importantly, it is shown that the model allows to tune the rate of coalescence in a wide range from fast to slow compared to bubble rearrangement dynamics.

\section{Pressure tensor}
\label{sec:PressureTensor}

In view of the paramount importance of pressure tensor in multiphase flows, we present here a standard derivation of this quantity within a free energy functional approach. The relation derived here will then be used in later section, where we consider a square gradient free energy model. Let $\mathcal{L}$ be a free energy density. The total free energy of the system is then given by \cite{Goldstein1980, Wheeler1996}

\begin{equation}
\mathcal{F} = \int_{\Omega}^{} \mathcal{L}(\phi, \nabla \phi, \vec{x})dV.
\label{Lagrangian}
\end{equation}

In Eq.~(\ref{Lagrangian}), $\mathcal{F}$ represents the action and $\mathcal{L}$ the Lagrangian free density in the defined control volume of $\Omega$. In general and to the first order in spatial gradients, $\mathcal{L}$ in Eq.~(\ref{Lagrangian}) depends on phase-field parameter, $\phi$, it's gradients, $\nabla \phi$, and the coordinate vector $\vec{x}$. Applying the variational principle to Eq.~(\ref{Lagrangian}), one obtains the well-known Euler-Lagrange equation,

\begin{equation}
\frac{\partial \mathcal{L}}{\partial \phi} - \frac{d}{dx_{i}} \frac{\partial \mathcal{L}}{\partial \partial_{i} \phi}= 0,
\label{Euler-Lagrange}
\end{equation}

where we used coordinate space $\vec{x} =(x_{1},x_{2},x_{3})$, Einstein summation convention, $a_{i}b_{i} =a_{1}b_{1}+a_{2}b_{2}+a_{3}b_{3}$, and $\partial_{i} \equiv \partial /\partial x_{i}$. Equation.~(\ref{Euler-Lagrange}) gives the equilibrium state of the system. In order to obtain an expression for the pressure tensor, one makes use of the connection between momentum conservation and symmetry properties of $\mathcal{L}$ \cite{Goldstein1980}. Keeping in mind that the translational invariance of $\mathcal{L}$ is a key issue here, we evaluate, 

\begin{equation}
\frac{d \mathcal{L}}{d x_{j}} =  \frac{\partial \mathcal{L}}{\partial \partial_{i} \phi} \partial_{j}  \partial_{i} \phi + \frac{\partial \mathcal{L}}{\partial \phi}  \partial_{j}  \phi + \frac{\partial
\mathcal{L}}{\partial x_{j}}.
\label{Tot-derivative-comp}
\end{equation}

To proceed further, $\partial \mathcal{L}/\partial \phi$ in Eq.~(\ref{Tot-derivative-comp}) is replaced by the second term in Eq.~(\ref{Euler-Lagrange}). This gives

\begin{eqnarray}
\frac{d \mathcal{L}}{dx_{j}}& =&  \frac{\partial \mathcal{L}}{\partial \partial_{i} \phi} \partial_{j}  \partial_{i} \phi + \Big( \frac{d}{dx_{i}} \frac{\partial \mathcal{L}}{\partial \partial_{i} \phi} \Big) \partial_{j}  \phi + \frac{\partial
\mathcal{L}}{\partial x_{j}} \nonumber \\
&=& \frac{d}{dx_{i}} \Big( \frac{\partial \mathcal{L}}{\partial \partial_{i} \phi} \partial_{j} \phi \Big) + \frac{\partial
\mathcal{L}}{\partial x_{j}},
\label{Tot-derivative-comp2}
\end{eqnarray}

where in the last line we applied the product rule. Finally, by rearranging the terms in Eq.~(\ref{Tot-derivative-comp2}) one obtains

\begin{equation}
-\frac{\partial\mathcal{L}}{\partial x_{j}} = \frac{d}{dx_{i}} \Big( -  \mathcal{L} \delta_{ij} + \frac{\partial \mathcal{L}}{\partial \partial_{i} \phi} \partial_{j} \phi \Big), \label{Tot-derivative-comp3}
\end{equation}

where $\delta_{ij}$ is the Kronecker delta. This is the key point to obtain the interfacial force. While the right hand side of Eq.~(\ref{Tot-derivative-comp3}) is the divergence of a tensor, the left hand side gives the variation of $\mathcal{L}$ with respect to $\vec{x}$. Assuming that the Lagrange density $\mathcal{L}$ does not explicitly depend on $\vec{x}$, i.e. $\mathcal{L}=\mathcal{L}(\phi, \nabla \phi)$, then the left hand side of Eq.~(\ref{Tot-derivative-comp3}) vanishes. This means that, in this case, there is a divergence free tensor (here called a pressure tensor), $\nabla \cdot \vec{P} = 0$, given by

\begin{equation}
\vec{P} = -  \mathcal{L} \vec{I} + \frac{\partial \mathcal{L}}{\partial \nabla \phi} \nabla \phi.
\label{PressTensV}
\end{equation}

In Eq.~(\ref{PressTensV}), $\vec{I}$ is the unit tensor and the second term is to be understood as a tensorial or dyadic product. It is important to note that the divergence of the pressure tensor is only zero at equilibrium. Beyond equilibrium, $\nabla \cdot \vec{P}$ plays a major role for the interface dynamics. The whole derivation until now was for a general Lagrangian density $\mathcal{L}$ which explicitly depends only on $\phi$ and its gradient $\nabla \phi$. In the next section, a specific form of Lagrangian density, which is called square gradient model, is employed to derive the exact form of the pressure tensor for a system of two immiscible phases. 

\section{A square gradient model} 
\label{sec:Squ}

\begin{figure}
\centering
\includegraphics[width=0.28\linewidth]{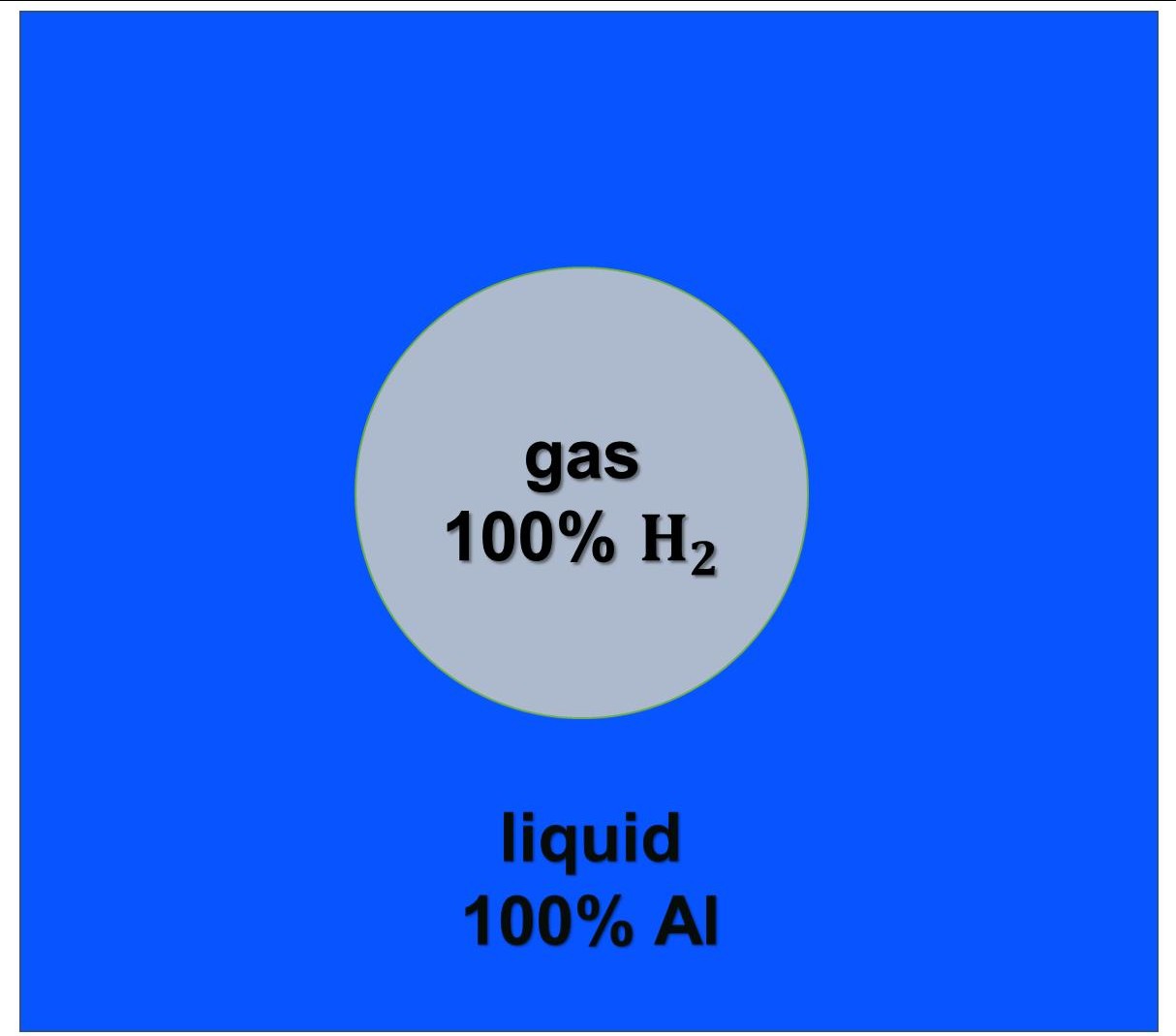}
\caption{Schematic plot of a two-phase system including a hydrogen (H$_{2}$) bubble surrounded by aluminum (Al) melt. In the current work, bubble is labeled as gas phase and the melt as liquid phase.}
\label{fig:Sch_bubble}
\end{figure}

Here, we define an idealized system of two completely immiscible phases, where each phase contains a different component. For instance, one can consider hydrogen (H$_{2}$) bubbles immersed in pure aluminum (Al) melt and assume that H$_{2}$ and Al do not mix. For simplicity, we only consider a single bubble immersed inside the melt (Fig.~\ref{fig:Sch_bubble}). Each phase is recognized by a phase-field parameter; $\phi_{\textrm{g}}$ for the bubble (g$=$gas) and $\phi_{\textrm{l}}$ for the melt (l$=$liquid). Following the convention in the multiphase-field approach \cite{Steinbach2009}, we identify $\phi_{\alpha}$ ($\alpha=$g,l) with the fraction of the volume element, $dV=d^3x$, occupied by the phase $\alpha$: $\phi_{\alpha}=dV_{\alpha}/dV \in [0,1]$. With this convention, $\phi_\textrm{g}+\phi_\textrm{l}=1$. Denoting for simplicity $\phi_\textrm{g}=\phi$, the phase-field associated with the liquid phase is readily obtained $\phi_\textrm{l}=1-\phi$. Therefore, inside the bubble $\phi=1$, outside of it $\phi=0$ and in the interface between bubble and melt $0<\phi<1$. With this convention, the square gradient model reads,

\begin{equation}
\mathcal{L} = \frac{W^2}{2} |\nabla \phi|^2 + \frac{\gamma}{4} \phi^2 (1-\phi)^2 + \Big(h(\phi) f_{\textrm{g}}(\rho_{\textrm{g}}) + \big(1-h(\phi)\big) f_{\textrm{l}}(\rho_{\textrm{l}})\Big),
\label{Square_gradient_model}
\end{equation}

where $W$, $\gamma$, $f_{\textrm{g}}$, and $f_{\textrm{l}}$ are the interface coefficient, a parameter to tailor the magnitude of double well potential, and the bulk free energy density of gas and liquid phases, respectively. Furthermore, $h(\phi)$ is the so-called interpolation function and has the same value as $\phi$ in the bulk. To be more specific, we assume that $h(1)=1$ and $h(0)=0$, corresponding to bulk gas and liquid phases, respectively. In the interface, $h$ is not necessarily identical to $\phi$, nevertheless, it is a continuous function and has a similar trend as $\phi$, $0<h(\phi)<1$. The first term in Eq.~(\ref{Square_gradient_model}) is the well-known square gradient term, and the second term is the double well potential. These terms account for the interface free energy density, while the last term inside brackets accounts for the bulk free energy density. For the specific case of $\mathcal{L}$ in Eq.~(\ref{Square_gradient_model}), one can determine the pressure tensor by substituting Eq.~(\ref{Square_gradient_model}) into Eq.~(\ref{PressTensV}). This yields

\begin{equation}
\vec{P} = -\bigg(  h(\phi) f_{\textrm{g}} + \big(1-h(\phi)\big) f_{\textrm{l}} + \frac{\gamma}{4} \phi^2 (1-\phi)^2 + \frac{W^2}{2} |\nabla \phi|^2 \bigg) \vec{I} + W^2 \nabla \phi\nabla \phi,
\label{Pressure_tensor_squareGM}
\end{equation}

where the last term is a dyadic product. Here, we introduce unit normal to the interface, $\vec{n} = -\nabla \phi/|\nabla \phi|$. Then, substituting $\nabla \phi = -\vec{n} |\nabla \phi|$ into Eq.~(\ref{Pressure_tensor_squareGM}), the pressure tensor can be reformulated as

\begin{equation}
\vec{P} = p_{\textrm{T}}(\vec{I} - \vec{n} \vec{n}) + p_{\textrm{N}} \vec{n} \vec{n} = p_{\textrm{T}} \vec{I} + ( p_{\textrm{N}}- p_{\textrm{T}})\vec{n} \vec{n},
\label{Pressure_tensor_components_general}
\end{equation}

where $\vec{n} \vec{n}$ is a dyadic product (and thus a tensor) and $p_{\textrm{N}}$ and $p_{\textrm{T}}$ are normal and tangential components of the pressure tensor, respectively, given by

\begin{eqnarray}
p_{\textrm{N}} &=&  -h(\phi) f_{\textrm{g}} - \big(1-h(\phi)\big) f_{\textrm{l}} - \frac{\gamma}{4} \phi^2 (1-\phi)^2 + \frac{W^2}{2} |\nabla \phi|^2, \nonumber\\
p_{\textrm{T}} &=&  -h(\phi) f_{\textrm{g}} - \big(1-h(\phi)\big) f_{\textrm{l}} - \frac{\gamma}{4} \phi^2 (1-\phi)^2 - \frac{W^2}{2} |\nabla \phi|^2.
\label{Normal_Tangential_general}
\end{eqnarray}

Beyond the interface, $p_{\textrm{N}} = p_{\textrm{T}} = p$, where $p$ is the bulk pressure. This is also seen from Eq.~(\ref{Normal_Tangential_general}), where the only difference between $p_{\textrm{N}}$ and $p_{\textrm{T}}$ is the last term, $ (W^2/2)(\partial \phi/\partial r)^2$, which is non-zero only in the interface region. Hence, $p_{\textrm{N}}-p_{\textrm{T}}$ is only non-zero in the interface. As will be shown later in Sec.~(\ref{subsec:Pla}), the integral of this quantity across the interface is identical to the interface free energy.\\
\\
To proceed further, the interfacial force is derived. Taking divergence of Eq.~(\ref{Pressure_tensor_components_general}) and reordering terms gives

\begin{equation}
\nabla \cdot \vec{P} = \nabla p_{\textrm{T}} + \Big(\nabla (p_{\textrm{N}} - p_{\textrm{T}} )\Big)\cdot \vec{n} \vec{n} + (p_{\textrm{N}} - p_{\textrm{T}}) \nabla \cdot (\vec{n} \vec{n}).
\label{Interfacial_force_NT}
\end{equation}

Then, we add zero (in the form of $\nabla p_{\textrm{N}}-\nabla p_{\textrm{N}} \cdot \vec{I}$) to the right hand side of Eq.~(\ref{Interfacial_force_NT}) to obtain

\begin{eqnarray}
\nabla \cdot \vec{P} &=& \nabla p_{\textrm{N}} + (p_{\textrm{N}} - p_{\textrm{T}}) \vec{n} \nabla \cdot \vec{n}  - (\vec{I} - \vec{n} \vec{n}) \cdot  \nabla ( p_{\textrm{N}} - p_{\textrm{T}} ) \nonumber\\
&=& \nabla p_{\textrm{N}} + (p_{\textrm{N}} - p_{\textrm{T}}) \kappa \vec{n},
\label{Interfacial_force_NT2}
\end{eqnarray}

where in the first line, the last term cancels out since $\nabla ( p_{\textrm{N}} - p_{\textrm{T}} ) $ is normal to the interface so that its scalar product with the projection operator $(\vec{I}-\vec{n}\vec{n})$ becomes zero. In the second line, $\kappa=\nabla \cdot \vec{n}$ is the mean curvature. A main advantage of deriving interfacial force in the form of Eq.~(\ref{Interfacial_force_NT2}) is its clear representation of the two major contributions from the interface; $\nabla p_{\textrm{N}}$ represents hydrostatic pressure force while $(p_{\textrm{N}} - p_{\textrm{T}}) \kappa \vec{n}$ corresponds to the curvature-induced force. This will be expanded in more details in Secs.~(\ref{subsec:Pla}) and~(\ref{subsec:Sta_bubble}) to determine the characteristics of the model for planar and curved interfaces. At equilibrium, $\nabla \cdot \vec{P} = 0$. Hence, integrating Eq.~(\ref{Interfacial_force_NT2}) across the interface in the normal direction leads to the diffuse interface version of the Young-Laplace equation,

\begin{equation}
\Delta p_{\textrm{N}} = \int \kappa (p_{\textrm{N}} - p_{\textrm{T}})  dn,
\label{Young_Laplace_general}
\end{equation}

where $\Delta p_{\textrm{N}} = p_{\textrm{N,g}} - p_{\textrm{N,l}} = p_\textrm{g}-p_\textrm{l}$ is the difference of pressure in bulk gas and liquid phases. Meanwhile, one can also obtain the relation between the pressure and bulk free energy by assigning $\phi=1$ and $\phi=0$ in Eq.~(\ref{Normal_Tangential_general}), respectively, for the gas and the liquid phases (recalling that $h(1)=1$ and $h(0)=0$). This gives,

\begin{eqnarray}\nonumber
p_{\textrm{g}} &=& -f_{\textrm{g}}\\
p_{\textrm{l}} &=& -f_{\textrm{l}}.
\label{Equation_of_state}
\end{eqnarray}

Thus, for the present model, the bulk free energy density of each phase is given by its equation of state (EOS). Note that, here, one needs to employ one EOS for each phase in the system. Some possible choices are ideal gas, Van der Waals or any other types of EOS.\\
\\
It will be shown in Sec.~(\ref{subsec:Pla}) that in the current model the interface energy, $\sigma$, is given by $\sigma = \int (p_{\textrm{N}} - p_{\textrm{T}})dn = \int W^2|\nabla \phi|^2 dn$. It is noteworthy that $\sigma = \int (p_{\textrm{N}} - p_{\textrm{T}})dn$ is a well known equation derived based on the mechanical equilibrium in the interface \cite{Rowlinson1982}. Thus, the right hand side of Eq.~(\ref{Young_Laplace_general}) can be approximated to $\sigma \kappa$ provided that $R\gg\eta$, where $R$ is the radius and $\eta$ is the width of the interface \cite{Vakili2017}. As a result, Eq.~(\ref{Young_Laplace_general}) is regarded as the diffuse-interface formulation of Young-Laplace equation, $\Delta p = \kappa \sigma$.\\
\\
The final form of the interfacial force can be obtained either by substituting Eq.~(\ref{Normal_Tangential_general}) into Eq.~(\ref{Interfacial_force_NT2}) or directly from the divergence of the pressure tensor in Eq.~(\ref{Pressure_tensor_squareGM}). In either case, it yields 

\begin{equation}
\nabla \cdot \vec{P} = \Big( (p_{\textrm{g}} - p_{\textrm{l}})\frac{\partial h}{\partial \phi} - \gamma \phi (1-\phi)(\frac{1}{2}-\phi) + W^2 \nabla^2 \phi \Big) \nabla \phi,
\label{Interfacial_force}
\end{equation}

where we also used Eq.~(\ref{Equation_of_state}). A similar equation is also presented in \cite{Gurtin1996, Chella1996} and is referred to as capillary force. The main difference of the interfacial force in Eq.~(\ref{Interfacial_force}) with those in  \cite{Gurtin1996, Chella1996} is that, in our approach, the interface energy in Eq.~(\ref{Interfacial_force}) ( $-\gamma \phi (1-\phi)(1/2-\phi) + W^2 \nabla^2 \phi$) is independent of the bulk free energy contribution ( $-(p_{\textrm{g}} - p_{\textrm{l}})(\partial h/\partial \phi)$). This feature will be clearly seen in Sec.~(\ref{sec:Model}) after determining the unknowns ($h(\phi)$, $\gamma$, $W$) in Eq.~(\ref{Interfacial_force}). In the following, firstly, the equilibrium condition for a planar interface is considered. Then, a single bubble in equilibrium with its surrounding liquid phase is studied. Through this analysis, the physical meaning of the model parameters is elucidated.

\subsection{Planar interface}
\label{subsec:Pla}

\begin{figure}
\centering
\includegraphics[width=0.28\linewidth]{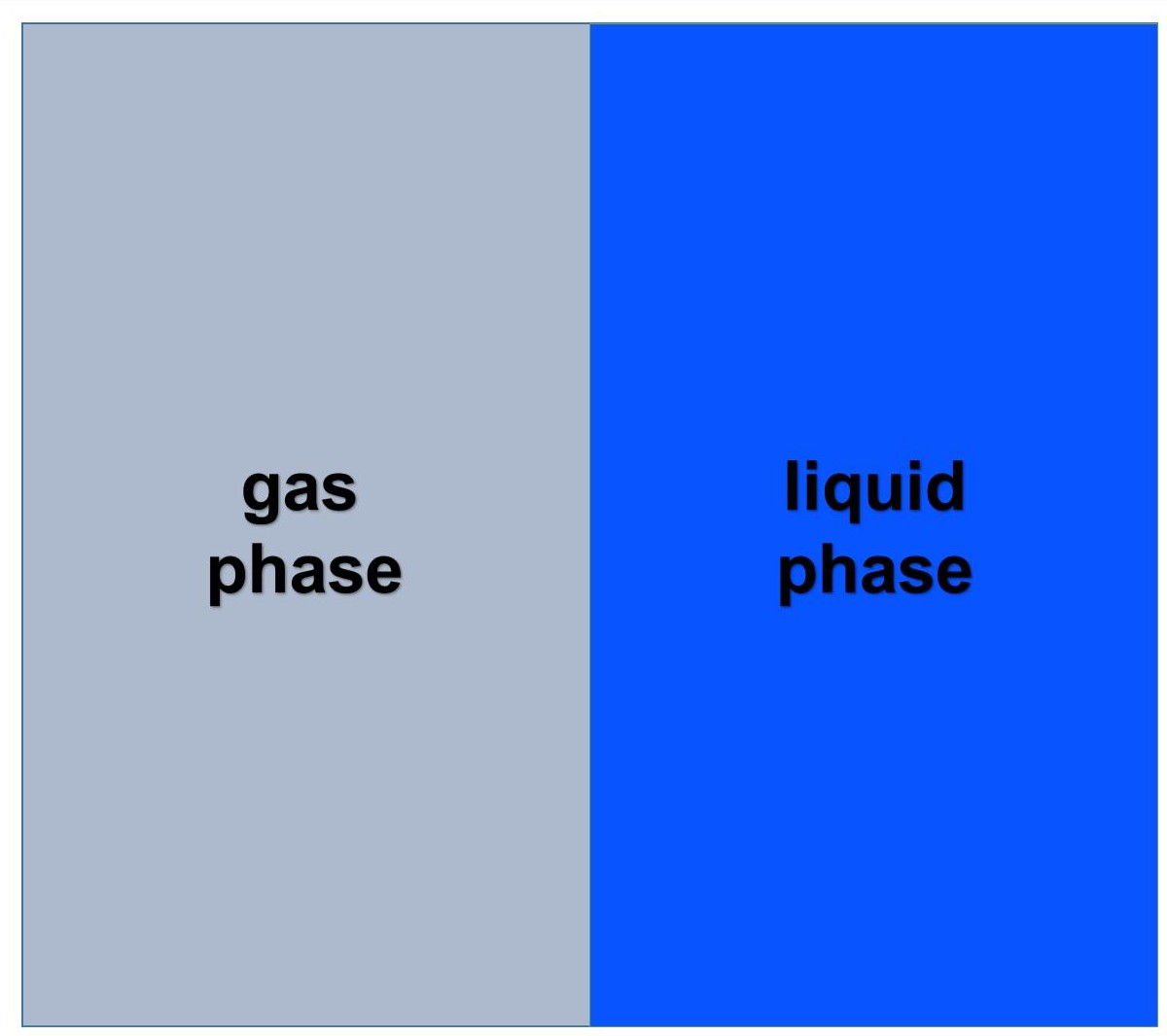}
\caption{Schematic view of a planar interface between gas and liquid phases in the equilibrium state in two dimensions. The center of the interface is at $x=0$. The system is homogenous along the vertical ($y$) direction.}
\label{fig:Planar}
\end{figure}

Although the study of planar interface is the simplest case, it is an essential step to derive the relation between model parameters, $W$, $\gamma$ and $\sigma$. To do so, we consider a system of two phases in two dimensions separated by a stable planar interface, Fig.~\ref{fig:Planar}. The interface is located at $x=0$ and spans in the $y$ direction. Since $\phi$ only varies in one direction ($x$), one obtains that $\vec{n}$ is the unit vector along the $x$-direction, so that its divergence vanishes. Thus, as expected for a planar interface, $\kappa$ vanishes and Eq.~(\ref{Young_Laplace_general}) turns into $\Delta p = 0$. This yields that the values of pressure in both phases are identical, $p_{\textrm{g}} = p_{\textrm{l}}$. Substituting this into Eq.~(\ref{Interfacial_force}), the first term cancels out and in the equilibrium, $\nabla \cdot \vec{P} = 0$, it yields

\begin{equation}
- \gamma \phi (1-\phi)(\frac{1}{2}-\phi) + W^2 \frac{\partial^2 \phi}{\partial x^2} = 0.
\label{Force_balance_palanar}
\end{equation}

It is easily verified that Eq.~(\ref{Force_balance_palanar}) is solved by the ansatz,

\begin{equation}
\phi(x) =\frac{1}{2}-\frac{1}{2}\tanh\big( \frac{6x}{\eta} \big),
\label{Interface_profile_planar}
\end{equation}

where $\eta$ is a parameter which determines the interface width. Substitution of Eq.~(\ref{Interface_profile_planar}) into Eq.~(\ref{Force_balance_palanar}) gives a first relation between the parameters of the square gradient free energy model on the one hand and $\eta$, on the other hand, 

\begin{equation}
\gamma = W^2\frac{288}{\eta^2}.
\label{Parameters}
\end{equation}

To proceed further, we evaluate the interface free energy per unit area by subtracting from Eq.~(\ref{Square_gradient_model}) the contribution associated with homogeneous phases essentially in the same spirit as in \cite{Cahn1958}. This gives, after integration over "volume" and dividing by surface area (a line segment in the 2D case considered here),

\begin{equation}
\sigma = \int_{-\infty}^{\infty} \bigg( \frac{W^2}{2}\Big(\frac{\partial \phi}{\partial x}\Big)^2+ \frac{\gamma}{4} \phi^2(1-\phi)^2 \bigg) dx.
\label{Interface_energy}
\end{equation}  

In addition, integrating Eq.~(\ref{Force_balance_palanar}) and using the fact that boundary terms vanish, one obtains $(W^2/2)(\partial \phi/\partial x)^2 = (\gamma/4)\phi^2(1-\phi)^2$. Inserting this result into  Eq.~(\ref{Interface_energy}), one arrives at,

\begin{equation}
\sigma =\int_{-\infty}^{\infty} W^2\Big(\frac{\partial \phi}{\partial x}\Big)^2 dx.
\label{Interface_energy2}
\end{equation}

From Eq.~(\ref{Normal_Tangential_general}), it is readily seen that the integrand of Eq.~(\ref{Interface_energy2}) is identical to $p_{\textrm{N}} - p_{\textrm{T}}$. Hence, Eq.~(\ref{Interface_energy2}) is in agreement with the mechanical definition of the interface energy, $\sigma = \int_{-\infty}^{\infty} (p_{\textrm{N}} - p_{\textrm{T}}) dx$ \cite{Rowlinson1982, Varnik2000}. Using the tanh ansatz for $\phi$, Eq.~(\ref{Interface_profile_planar}), the interface energy is calculated from Eq.~(\ref{Interface_energy2}). One thus obtains,
 
\begin{equation}
\sigma = -\int_{-\infty}^{\infty} W^2\Big( \frac{\partial \phi}{\partial x} \Big)^2 \frac{dx}{d \phi}  d \phi = \int_{0}^{1} \frac{12W^2}{\eta} \phi(1-\phi) d \phi = \frac{2W^2}{\eta}.
\label{Interface_energy3}
\end{equation}

Combining Eqs.~(\ref{Parameters}) and~(\ref{Interface_energy3}), $\gamma$ and $W^2$ are expressed in terms of interface free energy and width, $\sigma$ and $\eta$,

\begin{eqnarray}\nonumber
&\gamma &= \frac{144}{\eta} \sigma,\\
& W^2 &= \frac{\eta}{2} \sigma.
\label{Parameters2}
\end{eqnarray}

By this, most of the unknowns in Eq.~(\ref{Interfacial_force}) are determined except the interpolation function $h$. This will be discussed in Sec.~(\ref{subsec:Int_h}). However, prior to this, it is essential to evaluate the model for the case of a bubble in equilibrium with the surrounding liquid phase and derive the corresponding interfacial force balance, Sec.~(\ref{subsec:Sta_bubble}). This will prove useful also later, when we discuss the possible choices of the function $h(\phi)$.

\subsection{Stable bubble}
\label{subsec:Sta_bubble}

It is convenient to take the advantage of polar symmetry in the present case of a 2D gas bubble embedded in a liquid. For this reason, we let the phase-field profile $\phi \equiv \phi(r)$ to be a function of $r$, the distance from the origin of the bubble. It follows from radial symmetry of the problem that  $\nabla \phi = (\partial \phi/\partial r) \vec{\hat{e}_{r}}$, where $\vec{\hat{e}_{r}} = \vec{r}/r$ is the unit radial vector. Recalling that $(\partial \phi/\partial r)<0$ in the present setup, the unit vector normal to the interface is given by $\vec{n} = -\nabla \phi/|\nabla \phi| =  \vec{\hat{e}_{r}}$. Inserting this expression into Eq.~(\ref{Pressure_tensor_components_general}) and taking its divergence leads to

\begin{equation}
\nabla \cdot \vec{P} = \Big( \frac{\partial p_{\textrm{N}}}{\partial r} + \frac{1}{r}(p_{\textrm{N}}-p_{\textrm{T}}) \Big) \hat{\vec{e}}_{\vec{r}},
\label{Divergence_pressure_tensor}
\end{equation}

where $p_{\textrm{N}}$ and $p_{\textrm{T}}$ are given by (substituting $|\nabla \phi| = |\partial \phi/\partial r|$ in Eq.~(\ref{Normal_Tangential_general})),

\begin{eqnarray}
p_{\textrm{N}} &=& -h(\phi) f_{\textrm{g}} - \big(1-h(\phi)\big) f_{\textrm{l}} - \frac{\gamma}{4} \phi^2 (1-\phi)^2 + \frac{W^2}{2} \Big( \frac{\partial \phi}{\partial r} \Big)^2,\nonumber\\
p_{\textrm{T}} &=&  -h(\phi) f_{\textrm{g}} -\big(1-h(\phi)\big) f_{\textrm{l}} - \frac{\gamma}{4} \phi^2 (1-\phi)^2- \frac{W^2}{2} \Big( \frac{\partial \phi}{\partial r} \Big)^2.
\label{Normal_parallel}
\end{eqnarray}

In Eq.~(\ref{Divergence_pressure_tensor}), the first term accounts for the force arising from radial variations of the hydrostatic pressure and the second term is the surface tension force. Imposing the equilibrium condition, $\nabla \cdot \vec{P} = 0$, and integrating Eq.~(\ref{Divergence_pressure_tensor}) across the interface lead to

\begin{equation}
\Delta p_{\textrm{N}} = \int_{0}^{\infty} \frac{1}{r}(p_{\textrm{N}}-p_{\textrm{T}})dr,
\label{Young_Laplace_bubble}
\end{equation}

where $r=0$ corresponds to the center of the bubble and $r=\infty$ to the liquid phase. Accordingly, $\Delta p_{\textrm{N}} \equiv p_{\textrm{N}}(0) -p_{\textrm{N}}(\infty) = p_{\textrm{g}}-p_{\textrm{l}}$ is the well known Laplace-pressure. Equation~(\ref{Young_Laplace_bubble}) is the special form of Eq.~(\ref{Young_Laplace_general}) for the case of a single bubble assuming its radial symmetry. In Eq.~(\ref{Young_Laplace_general}), $\kappa$ is obtained from its general definition ($\kappa =\nabla \cdot \vec{n}$) while $1/r$ in Eq.~(\ref{Young_Laplace_bubble}) emerges from radial symmetry of the phase-field, $\phi(r)$. Strictly speaking, the polar variable $r$ is not the radius of curvature of the bubble. This creates a discrepancy because the mean curvature of the bubble in two dimensions is defined as inverse of its radius, $\kappa=1/R$, which is a constant quantity. Thus, only in the center of the interface, $r=R$, one can assure that $\kappa = 1/r$ is satisfied. However, recalling that $R-\eta/2 \leq r \leq R+\eta/2$, it is easily seen that $r \approx R$ if $R \gg \eta/2$. Thus, on the right hand side of Eq.~(\ref{Young_Laplace_bubble}), $1/r$ can be approximated by $1/R$,  where $R$ is the bubble radius, provided that $R \gg \eta/2$ or simply $R\gg\eta$. Therefore, in this limit, Eq.~(\ref{Young_Laplace_bubble}) approximately reproduces the well-known Young-Laplace law.

\subsection{Interpolation function $h(\phi)$}
\label{subsec:Int_h}

The so-called interpolation function $h(\phi)$, as mentioned before, takes values of $0$ or $1$ in the bulk phases and varies in between in the interface, $0<h(\phi)<1$. It thus has a similar functionality as $\phi$. There are a number of alternatives for $h$ fulfilling this property such as $\phi$, $\phi^2(3-2\phi)$ and $\phi^3(6\phi^2-15\phi+10)$. However, here in this section, we will show that the choice of $h(\phi)$ is not completely arbitrary and it depends on the physical problem. To see this, we insert Eq.~(\ref{Normal_parallel}) into Eq.~(\ref{Divergence_pressure_tensor}) and use the force balance ($\nabla \cdot \vec{P} = 0$),

\begin{equation}
\Delta p \frac{\partial h(\phi)}{\partial \phi} - \gamma \phi (1-\phi)(\frac{1}{2}-\phi) + W^2 \frac{\partial^2 \phi}{\partial r^2} + \frac{W^2}{r} \frac{\partial \phi}{\partial r} = 0,
\label{Force_balance_bubble}
\end{equation}

where $\Delta p = p_{\textrm{g}}-p_{\textrm{l}}$ is the Laplace pressure and $p_{\textrm{g}} = -f_{\textrm{g}}$ and $p_{\textrm{l}}=-f_{\textrm{l}}$. Since there is only one equation, Eq.~(\ref{Force_balance_bubble}), and two unknowns, $h(\phi)$ and $\phi(r)$, then a simple way is to define one function, say $\phi$, and derive the other one, $h(\phi)$. A reasonable choice is $\phi(r) =0.5-0.5\tanh(6(r-R)/\eta)$, where $R$ is the radius of the bubble and $\eta$ is the interface width. As a result, $(\partial^2 \phi /\partial r^2) = (288/\eta^2)\phi(1-\phi)(0.5-\phi)$ and substituting this into Eq.~(\ref{Force_balance_bubble}) and using Eq.~(\ref{Parameters2}) gives

\begin{equation}
\Delta p \frac{\partial h(\phi)}{\partial \phi} = -\frac{\eta \sigma}{2r} \frac{\partial \phi}{\partial r}.
\label{Force_balance_bubble2}
\end{equation}

Equation~(\ref{Force_balance_bubble2}) resembles the Young-Laplace equation $\Delta p = \sigma \kappa$, where $1/r$ plays the role of curvature $\kappa$ (assuming $R \gg \eta$, Sec.~(\ref{subsec:Sta_bubble})), provided that
 
\begin{equation}
\frac{\partial h(\phi)}{\partial \phi} = -\frac{\eta}{2} \frac{\partial \phi}{\partial r}.
\label{Condition_Laplace_equation}
\end{equation}

Thus, using the fact that $\partial \phi/\partial r = -(12/\eta) \phi(1-\phi)$, Eq.~(\ref{Condition_Laplace_equation}) is integrated to yield

\begin{equation}
h(\phi) = \phi^2(3-2\phi).
\label{Interpolation_function}
\end{equation}

Noteworthy, this result is very similar to the function used in the phase-field method for solidification and grain growth \cite{Steinbach2009}. Finally, all the unknowns in Eq.~(\ref{Interfacial_force}) are determined and one can  use the interfacial force in the Navier-Stokes equations to compute the dynamics of the system. The governing equations of the current model are completely given in following section. However, prior to that, it is helpful to recast the free energy functional $\mathcal{F} = \int \mathcal{L}(\phi, \nabla \phi) d^3 \vec{x}$ by using Eqs.~(\ref{Square_gradient_model}),~(\ref{Equation_of_state}), and~(\ref{Parameters2}),

\begin{equation}
\mathcal{F} = \int_{\Omega}^{} \bigg( \sigma \Big( \frac{\eta}{4} |\nabla \phi|^2 + \frac{36}{\eta} \phi^2 (1-\phi)^2 \Big) -  \Big( h(\phi)p_{\textrm{g}} + \big(1-h(\phi)\big)p_{\textrm{l}} \Big) \bigg) d^3 \vec{x},
\label{Free_energy_functional}
\end{equation}

where the term multiplied with $\sigma$ accounts for the interface energy contribution while the rest gives the bulk free energy. This equation will be used below, when we address the fluid dynamic and the phase-field equations. 

\section{Model} 
\label{sec:Model}

For a system of two immiscible phases, the Navier-Stokes equations read

\begin{equation}
\rho \big( \frac{\partial \vec{u}}{\partial t} + \vec{u} \cdot \nabla \vec{u} \big) = -\nabla \cdot \vec{P} + \nabla \cdot \big( \mu (\nabla \vec{u} +\nabla \vec{u}^T) \big) + f^{\textrm{ext}},
\label{Navier_Stokes}
\end{equation}

where $\rho$, $\vec{u}$, $\mu$, and $f^{\textrm{ext}}$ correspond to density, velocity, viscosity, and the external force, respectively. In the present model, density is given by $\rho = \rho_{\textrm{g}} \phi +\rho_{\textrm{l}} (1-\phi)$, where $\rho_{\textrm{g}}$ and $\rho_{\textrm{l}}$ are the densities of gas and liquid phases, respectively. The mass of each phase is supposed to be constant. Thus, any variation in the volume of each phase leads to the density change of the same phase. Therefore, the densities are updated via

\begin{eqnarray}
\rho_{\textrm{g}} &=& \frac{M_{\textrm{g}}}{V_{\textrm{g}}},\nonumber \\
\rho_{\textrm{l}} &=&  \frac{M_{\textrm{l}}}{V_{\textrm{l}}},
\label{density}
\end{eqnarray}

where $M_{\textrm{g}}$ and $M_{\textrm{l}}$ are the (constant) masses of the gas and the liquid phases, respectively. The volumes of the corresponding phases are obtained from the integral of the phase-field parameter, $V_{\textrm{g}}=\int \phi dV$ and $V_{\textrm{l}} = \int (1-\phi) dV$. Furthermore, $\nabla \cdot \vec{P}$ in Eq.~(\ref{Navier_Stokes}) is the interfacial force, which is given by

\begin{equation}
\nabla \cdot \vec{P} = \bigg( 6\phi(1-\phi)\Delta p + \sigma \Big(- \frac{144}{\eta} \phi (1-\phi)(\frac{1}{2}-\phi) + \frac{\eta}{2} \nabla^2 \phi \Big) \bigg) \nabla \phi.
\label{Interfacial_force_model}
\end{equation}

Equation~(\ref{Interfacial_force_model}) is derived by substituting Eqs.~(\ref{Equation_of_state}),~(\ref{Parameters2}), and~(\ref{Interpolation_function}) into Eq.~(\ref{Interfacial_force}). Then, in order to obtain the interfacial dynamics in the presence of flow, we employ the standard phase-field ansatz,

\begin{equation}
\frac{\partial \phi}{\partial t}+ \vec{u} \cdot \nabla \phi = -\mathcal{M} \frac{\delta \mathcal{F}}{\delta \phi} = -\mathcal{M} \Big(\frac{\partial \mathcal{L}}{\partial \phi}-\nabla \cdot \frac{\partial \mathcal{L}}{\partial \nabla \phi}\Big),
\label{Phase_field_ansatz}
\end{equation}

where $\delta \mathcal{F}/\delta \phi$ is the functional derivative and $\mathcal{M}$ is the interface mobility. The exact form of the phase-field equation is obtained by inserting Eq.~(\ref{Free_energy_functional}) into Eq.~(\ref{Phase_field_ansatz}). Thus, it reads

\begin{equation}
\frac{\partial \phi}{\partial t}+ \vec{u} \cdot \nabla \phi = \mathcal{M} \bigg( 6\phi(1-\phi)\Delta p + \sigma \Big(- \frac{144}{\eta} \phi (1-\phi)(\frac{1}{2}-\phi) + \frac{\eta}{2} \nabla^2 \phi \Big) \bigg).
\label{Phase_field_equation}
\end{equation}

Equations~(\ref{Navier_Stokes}) and~(\ref{Phase_field_equation}) are the governing equations for the current model. The validity of the model is tested via a number of simulations described in the next section.

\section{Results and discussion}
\label{sec:Res}
\subsection{Simulation details}
\label{sec:Sim_det}

We use finite difference method to discretize the Navier-Stokes and phase-field equations, Eqs.~(\ref{Navier_Stokes}) and~(\ref{Phase_field_equation}). For the Navier-Stokes equations, we use forward time central space scheme in \cite{Abdallah1987}, except for the nonlinear velocity term which we apply upwind scheme. For the phase-field equation, we use forward in time and central scheme in space. Laplacian operator, $\nabla^2 \phi$, is discretized via a nine-points scheme \cite{Vakili2017}. Moreover, concerning $\Delta p = p_{\textrm{g}}-p_{\textrm{l}}$ in Eqs.~(\ref{Navier_Stokes}) and~(\ref{Phase_field_equation}), we account for ideal gas equation of state for the gas phase $p_{\textrm{g}} = c_{s,g}^{2}\rho_{\textrm{g}}$ and Van der waals equation of state for the liquid phase $p_{\textrm{l}} = (a \rho_{\textrm{l}}/(b-\rho_{\textrm{l}})-c \rho_{\textrm{l}}^2 $ , where $c_{s,g}$ is the speed of sound in the gas phase and $a$, $b$, and $c$ are constants set to $a=6.4$, $b=3$ and $c=3$. Furthermore, as mentioned earlier in this paper, Sec.~(\ref{subsec:Int_h}), in virtue of keeping numerical error small, unless otherwise stated, a ratio of $R/\eta = 10$ is chosen for all simulations. For simplicity, viscosities of both gas and liquid phases are chosen to be equal $\mu = \mu_\textrm{g} = \mu_\textrm{l} = 1$.\\
\\
We perform three benchmark tests. The first one is concerned with the stability of a 2D bubble embedded in a liquid. For this case, the simulations are performed in three subcategories each with a different interface energy and for each interface energy we consider four different bubble radii. The initial radius of the bubble in half of the simulations is chosen to be larger than the expected equilibrium value in order to capture the shrinkage while in the other cases bubbles smaller than the equilibrium size are initialized to see whether the model also captures the expansion of the bubble until it reaches the static equilibrium. The second benchmark test is to simulate the coalescence of two bubbles to see if the final bubble recovers the circular shape. A question of central importance in modeling of structure formation in metallic foams regards a safe control of the rate of coalescence. To check this issue, a set of simulations is performed for three different interface energies, $\sigma_{1} = 18.2$ (in dimensionless units), $\sigma_{2} = 10^{-2} \sigma_{1}$, and $\sigma_{3} = 10^{-4} \sigma_{1}$. In order to explore this aspect further, a third benchmark test is performed, consisting of two coalescing bubbles in the presence of a rotational flow. In this third test, it is demonstrated that the model allows full control over the rate of coalescence in such a way that when two bubbles come into contact, they can rotate around each other with almost no progress in their merging process. 

\subsection{Stability of a single bubble}
\label{subsec:stability_bubble}
\begin{figure}
\centering
\includegraphics[width=0.65\linewidth]{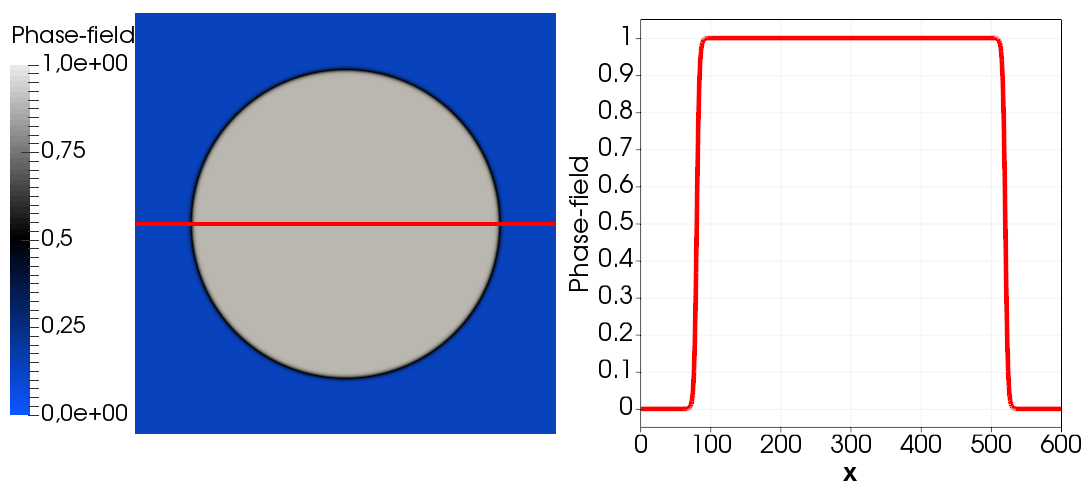}\\
(a) \hspace{40mm} (b)
\caption{(a) Final state of a 2D stable bubble in the equilibrium state. The color code corresponds to $\phi$. (b) The equilibrium profile of the phase-field along a line passing through the bubble center (the red line in(a)).}
\label{fig:Single_bubble}
\end{figure}

\begin{figure}
\centering
\includegraphics[width=0.4\linewidth]{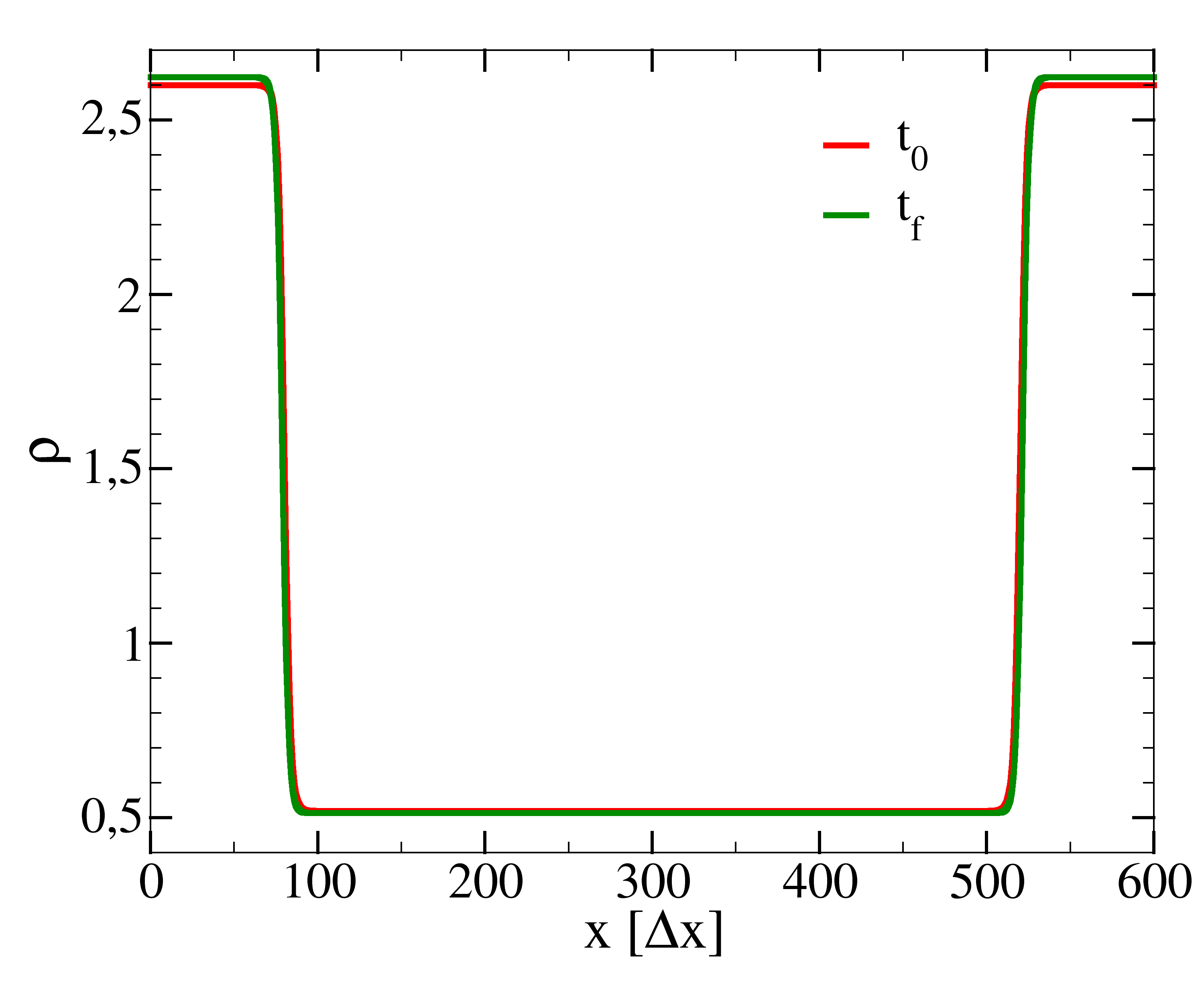}
\includegraphics[width=0.4\linewidth]{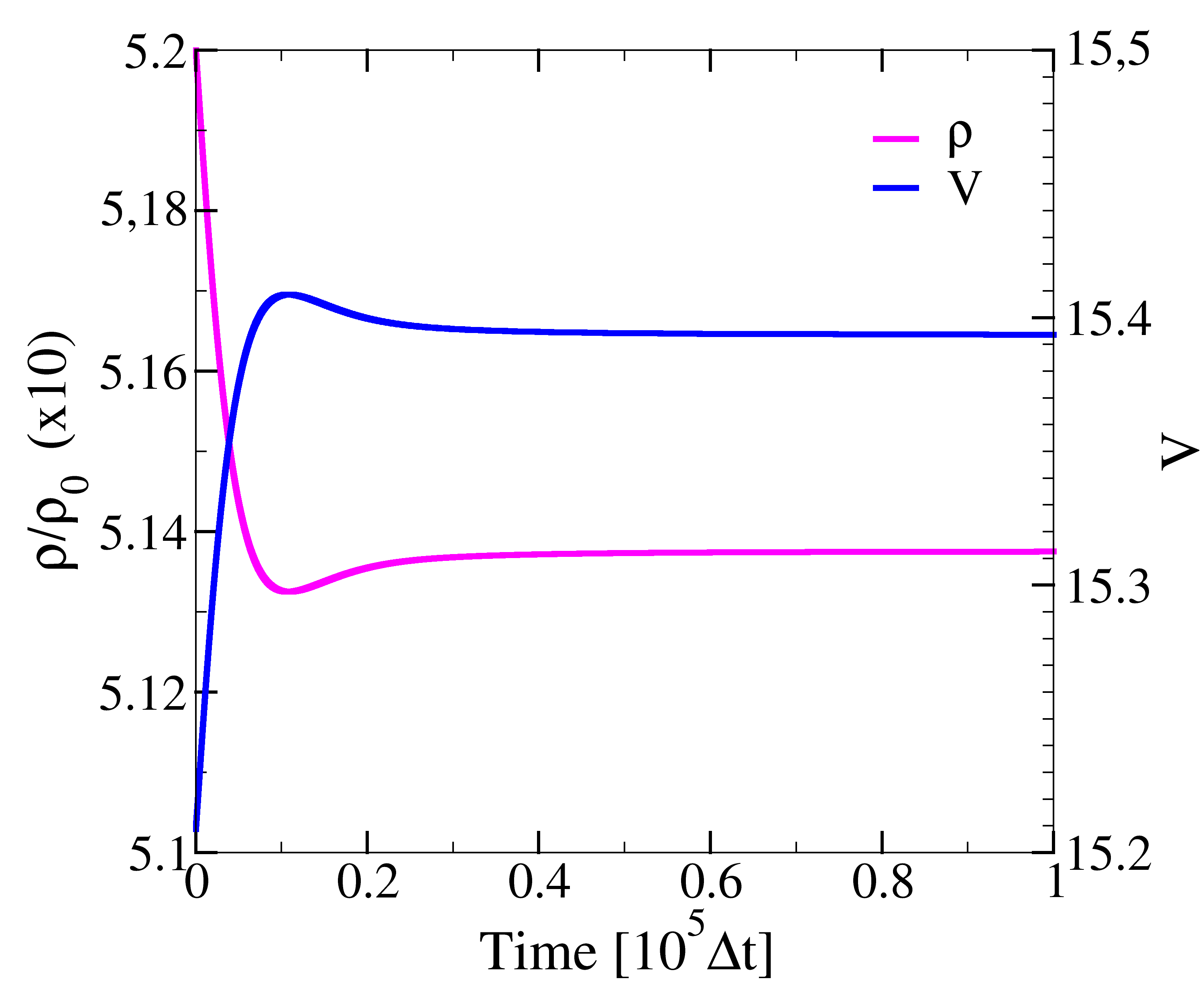}\\
(a) \hspace{40mm} (b)\\
\includegraphics[width=0.4\linewidth]{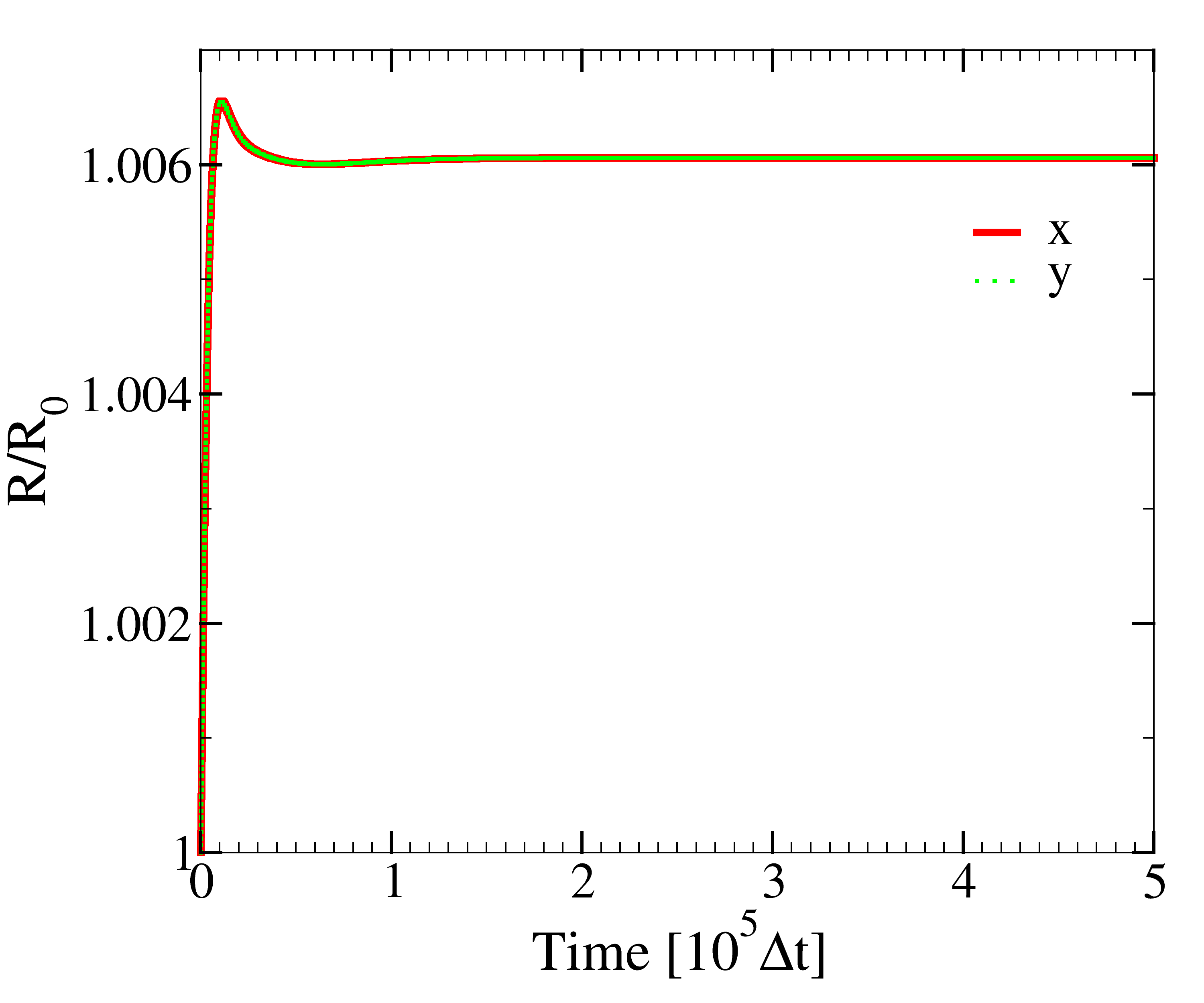}\\
(c)
\caption{(a) Density profile along the red line in Fig.~\ref{fig:Single_bubble}-a in the initial state ($t_{0}$) and final static equilibrium state ($t_\textrm{f}$). (b) Gas density and the corresponding bubble volume versus time, highlighting the decrease in gas density during the initial expansion process. Note that $\rho_0\equiv \rho(t=0)=1$. (c) Variation of bubble radius with time in the horizontal $x$ (solid line) and vertical $y$ (dashed line) directions.}
\label{fig:Radius}
\end{figure}

Figure~\ref{fig:Single_bubble} shows the result of a simulation for a static bubble in equilibrium. The phase-field parameter $\phi$ is indicated via color code (Fig.~\ref{fig:Single_bubble}-a) and its profile along the center line (red) is plotted in Fig.~\ref{fig:Single_bubble}-b. $\phi=1$ corresponds to the bubble/gas phase, $\phi=0$ to the liquid phase, and $0<\phi<1$ to the interface in between. Moreover, the contour line, defined via $\phi=0.5$, is supposed to represent the position of a "dividing surface" (here a circular line) between the two phases, where proportion of each phase is $50\%$ and $r=R$. Figure~\ref{fig:Radius}-a shows the density profiles across the center line in the initial $(t_{0})$ and final static equilibrium condition $(t_\textrm{f})$. A careful survey of the final equilibrium density profile at $t_\textrm{f}$ (Fig.~\ref{fig:Radius}-a) reveals increase of $\rho_\textrm{l}$ while $\rho_\textrm{g}$ decreases slightly. To better visualize the decrease of the gas density, we plot it separately as a function of simulation time in Fig.~\ref{fig:Radius}-b. Since the mass of each phase is considered to be constant, the variation of the density is caused by volume change only. This is clearly visible in Fig.~\ref{fig:Radius}-b where the increase of gas volume coincides with the decrease of its density. The opposite trend is also expected for the liquid phase as it shrinks and becomes denser. The plateau in Fig.~\ref{fig:Radius}-b corresponds to the static equilibrium and it retains this until the end of simulation. A nice feature of the present model is the possibility of defining appropriate equation of states and as a result adjusting the compressibility of each phase. By virtue of this property, one can cover a wide range of densities and compressibilities for the simulation of different materials.\\
\\
The current model is thus capable of restoring the expected static equilibrium, if the initial configuration deviates from it. For the same system, in order to show the variation of the bubble shape due to the volume change, we plot two perpendicular radii of the bubble versus time in Fig.~\ref{fig:Radius}-c. Here, we define the radius along a given direction as half of the distance between two intersecting points of the contour line $\phi=0.5$ with a straight line along that direction. Figure~\ref{fig:Radius}-c shows that these quantities, two perpendicular radii,  vary until they reach and maintain a constant value. During these simulations, both radii are always identical and thus the bubble maintains its circular shape during the entire simulation. This is expected, since in the present example of a single bubble, only isotropic forces are present.\\

\begin{figure}
\centering
\includegraphics[width=0.45\linewidth]{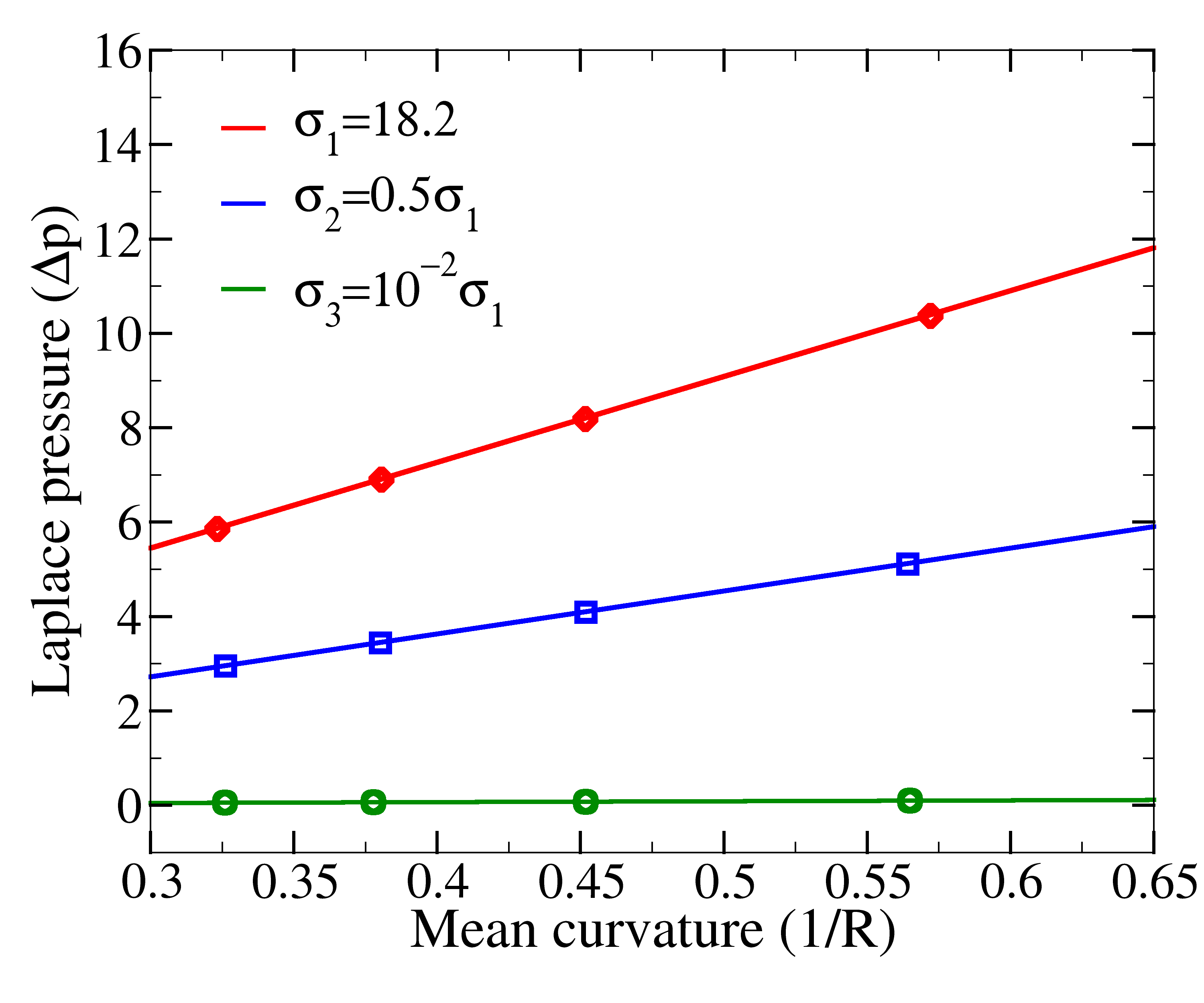}
\caption{Variation of the Laplace pressure ($\Delta p$) with the mean curvature ($1/R$). The symbols depict the simulation results and the solid lines are the analytical predictions, $\Delta p = \sigma/R$ using $\sigma$ from Eq.~(\ref{Interface_energy3}). The reference interface energy $\sigma_1$ is chosen to be $\sigma_1=18.2$.}
\label{fig:Laplace_law}
\end{figure}

Moreover, as shown in Fig.~\ref{fig:Laplace_law}, for the case of a single bubble, the variation of the Laplace pressure (pressure difference between the gas and the liquid phases) with mean curvature obeys the Young-Laplace equation,

\begin{equation}
\Delta p = p_{g} - p_{l} = \sigma \kappa,
\label{Young-Laplace_law}
\end{equation}

where $p_\textrm{g}$ and $p_\textrm{l}$ are gas and liquid pressures, respectively. Figure~\ref{fig:Laplace_law} depicts the simulation and analytical results for Laplace pressure versus mean curvature for three different interface energies $\sigma_{1} = 18.2$, $\sigma_{2}=0.5 \sigma_{1}$, and $\sigma_{3} = 10^{-2}\sigma_{1}$. The symbols indicate the result of the simulations while the solid lines represent $\Delta p= \sigma \kappa$ (a line with the slope of $\sigma$), where the interface energy $\sigma$ is an input parameter of the model. The simulation results are consistent with the analytical one by an error of less than $0.1 \%$. This reveals the capability of the present model to acquire the expected results for a wide range of interface energies, covering here two decades, $\sigma_{1}/\sigma_{3}=100$. Even higher ratios are also accessible. In the following section a ratio of $10^4$ is used to control the coalescence rate.

\subsection{Coalescence of bubbles}

\begin{figure}
\centering
$t = 0$\hspace{9mm}\includegraphics[width=0.6\linewidth]{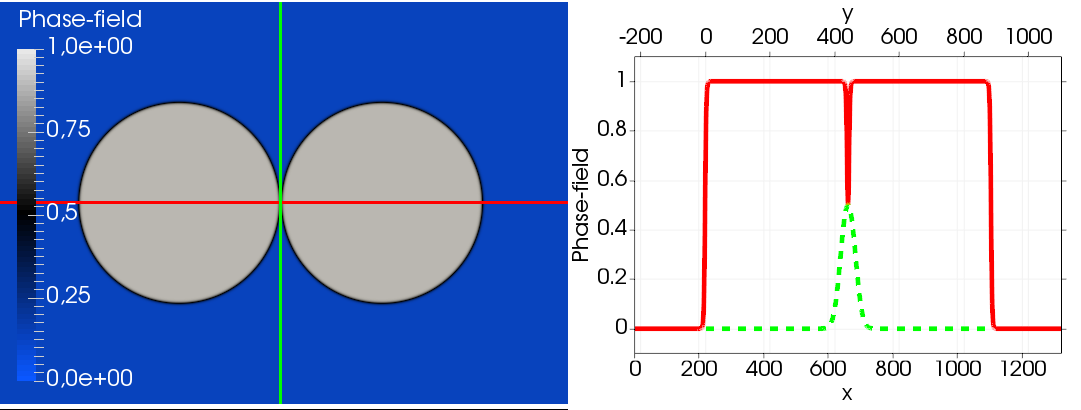}\\
$t = 2\times10^5$\includegraphics[width=0.6\linewidth]{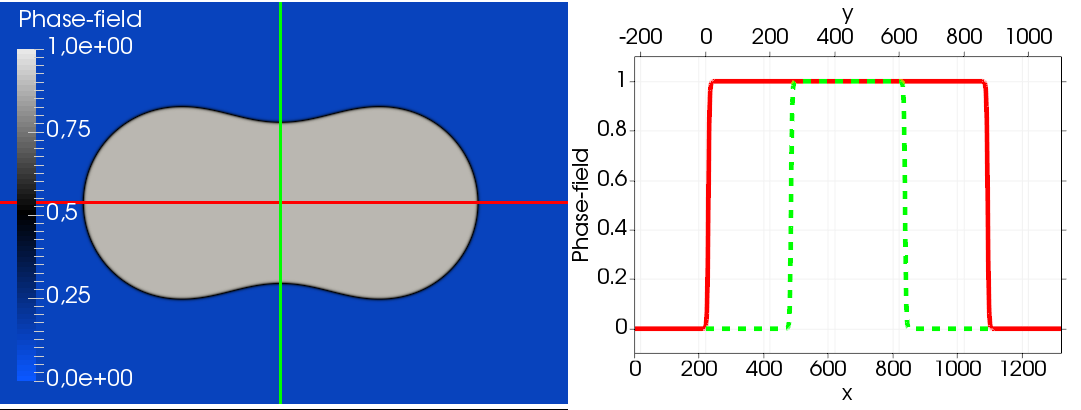}\\
$t = 5\times10^5$\includegraphics[width=0.6\linewidth]{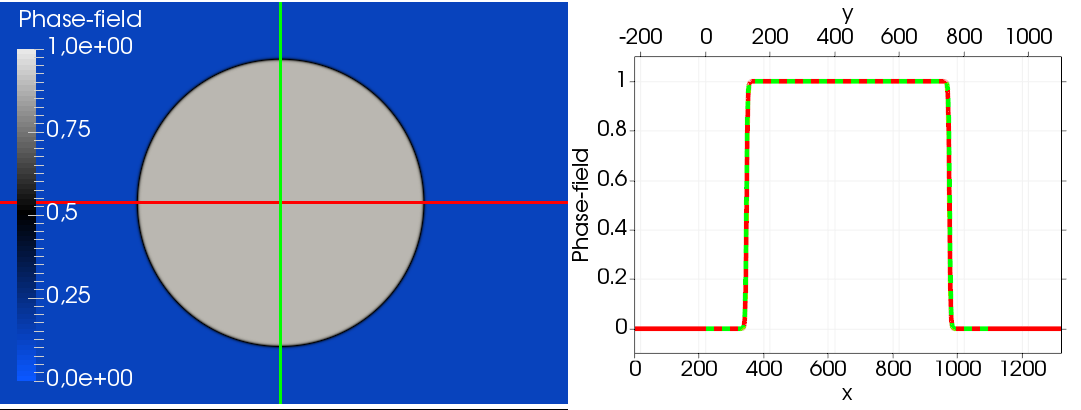}\\
\hspace{10mm }(a) \hspace{45mm} (b)
\caption{Time evolution of two coalescing bubbles until they completely merge into a single static bubble. The phase-field $\phi$ is indicated by the color code in (a) and its profile along the horizontal (red) and vertical (green) center lines is plotted in (b).}
\label{fig:Coalescence_stable}
\end{figure}

\begin{figure}
\centering
\includegraphics[width=0.45\linewidth]{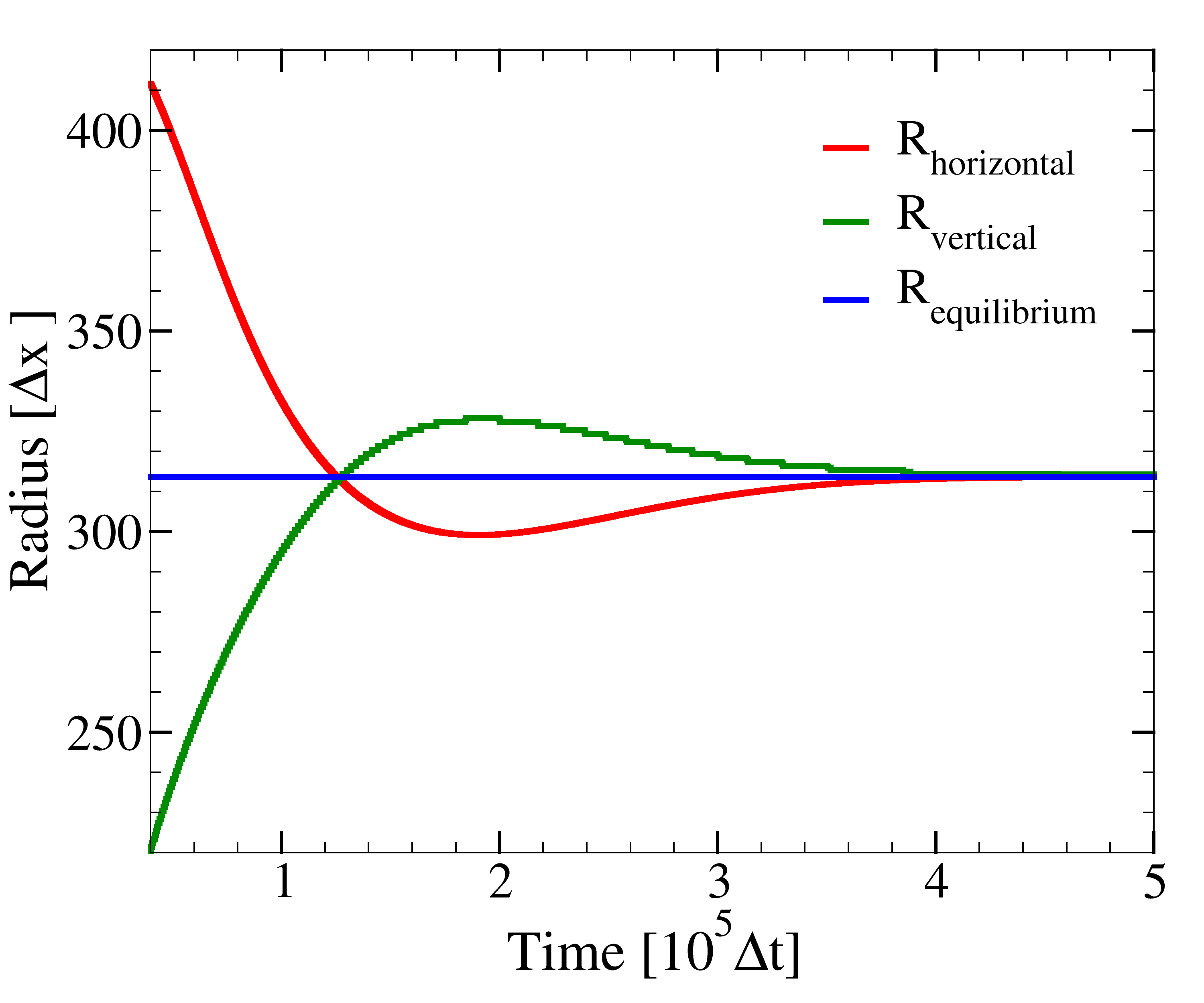}\\
\caption{Radii of the gas domain which forms out of two merging bubbles versus simulation time during the coalescence process. The red and green lines show the horizontal and vertical extensions, respectively, while the blue line shows the final equilibrium radius of the resulting single bubble.}
\label{fig:Coalescence_radii}
\end{figure}

The first benchmark test in the bubble coalescence is to check if two bubbles can merge completely and recover a circular shape. The result of this simulation is shown in Fig.~\ref{fig:Coalescence_stable} for three different time steps, initial, middle, and final (equilibrium). The phase-field ($\phi$) is represented by a color code in Fig.~\ref{fig:Coalescence_stable}-a and its profile along horizontal (red) and vertical center (green) lines is illustrated in Fig.~\ref{fig:Coalescence_stable}-b. The coalescence process advances until it recovers a single bubble (Fig.~\ref{fig:Coalescence_stable}-a). At this point, the $\phi$ profiles along the horizontal and vertical lines overlap, Fig.~\ref{fig:Coalescence_stable}-b. This confirms that the bubble recovers a circular shape. Furthermore, similar to previous section, the radius in horizontal and vertical directions can be calculated as half of the distance between intersection points of each line with contour line $\phi=0.5$. Variations of the horizontal and vertical radii in Fig.~\ref{fig:Coalescence_radii} show that they converge to the equilibrium radius and retain it.\\
\\

\begin{figure}
\centering
\includegraphics[width=\linewidth]{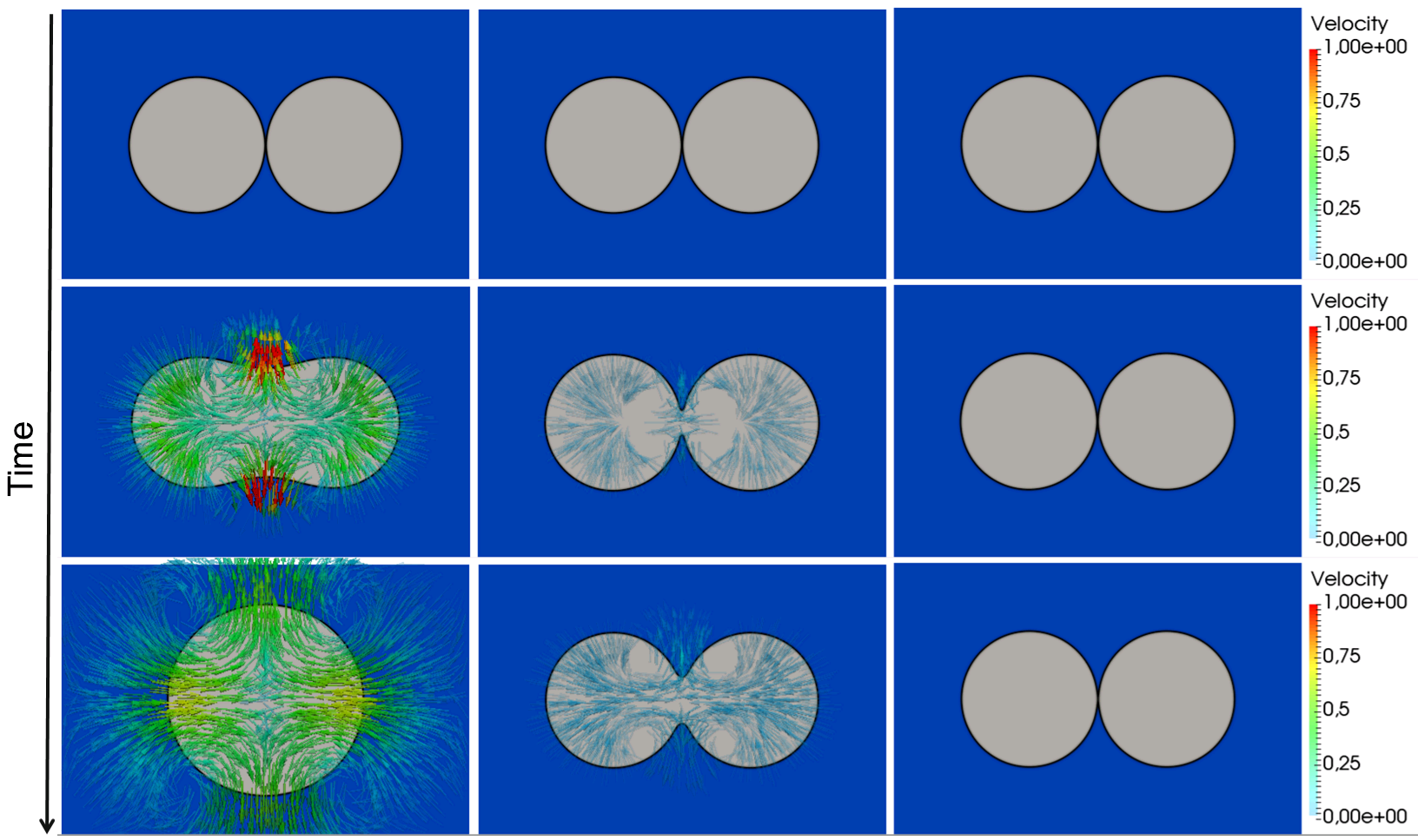}\\
\hspace{4mm}(a) \hspace{42mm} (b)  \hspace{42mm} (c) \hspace{10mm}
\caption{Time evolution of two coalescing bubbles with different interface energies, (a) $\sigma_{1} = 18.2$, (b) $\sigma_{2} = 10^{-2} \sigma_{1}$, and (c) $\sigma_{3}  = 10^{-4} \sigma_{1}$, where $\sigma_{3}=10^{-2}\sigma_{2}=10^{-4}\sigma_{1}$. The color code in the right corresponds to the magnitude of the velocity field.}
\label{fig:Coalescence_rate}
\end{figure}

The capability of the model to simulate the coalescence of bubbles leading to an equilibrium single bubble is already checked. The next step will be to decrease the coalescence rate. This is done here by reducing the interface energy between the gas and the liquid phases. Thus, a set of simulations is performed for coalescence of two bubbles with three different interface energies, $\sigma_{1}=18.2$, $\sigma_{2}=10^{-2}\sigma_{1}$ and $\sigma_{3}=10^{-4}\sigma_{1}$. The result of simulation for each interface energy is plotted for three different time steps in  Fig.~\ref{fig:Coalescence_rate}. In there, each row corresponds to the same time step starting from top to bottom, respectively, as initial to final configuration of two coalescing bubbles. For the case of $\sigma_{1}$, the coalescence of the bubbles leads to a circular single bubble at the final time step, Fig.~\ref{fig:Coalescence_rate}-a, while at the same time step for the case of $\sigma_{2}$, the coalescence process advances only partially, Fig.~\ref{fig:Coalescence_rate}-b. This is also confirmed from the data of the velocity field. The coalescence rate of the simulation with $\sigma_{2}$ is lower due to the smaller magnitude of the interface velocity compared to that of $\sigma_{1}$, Fig.~\ref{fig:Coalescence_rate}-a and~b. For simulation with the lowest interface energy, $\sigma_{3} = 10^{-4} \sigma_{1}$, the coalescence rate is so low that no interface velocity is visible within defined data range for the velocity field. Thus, the coalescence is almost suppressed and the bubbles keep their initial configuration during the simulated time window, Fig.~\ref{fig:Coalescence_rate}-c.\\
\\

\begin{figure}
\centering
\includegraphics[width=0.8\linewidth]{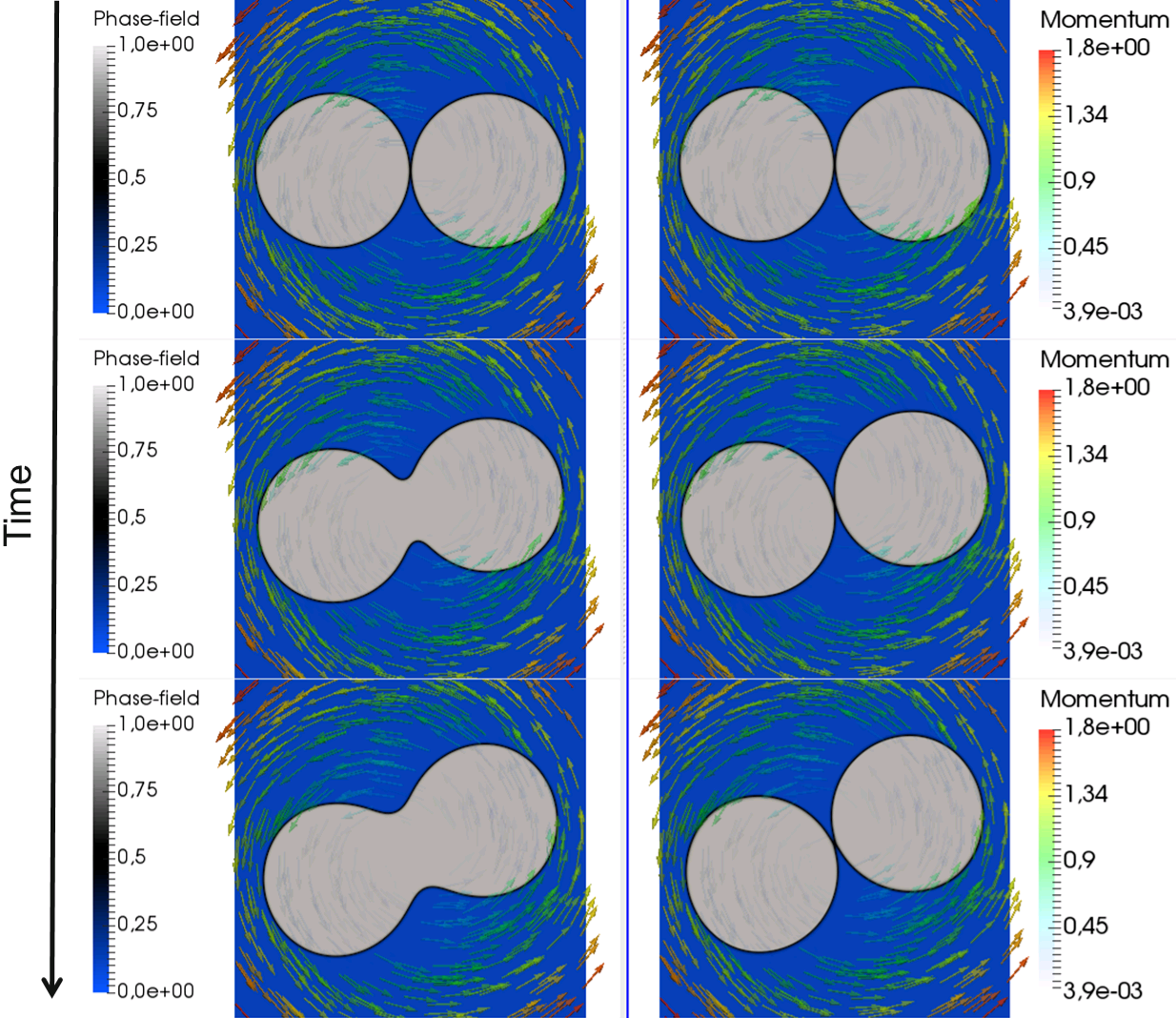}\\
(a) \hspace{40mm} (b)
\caption{Simulation of two coalescing bubbles in the presence of rotational flow for two different interface energies, (a) $\sigma_{2} = 10^{-2} \sigma_{1}$ and (b)  $\sigma_{3} =10^{-2}\sigma_{2}= 10^{-4} \sigma_{1}$. The figure in top corresponds to the initial simulation setup, while the figures in the middle and bottom correspond to middle and final time steps, respectively, until two bubbles rotate for $\pi/4$ radian. The color code on the left accounts for the phase-field parameter, while that in the right accounts for velocity magnitude.}
\label{fig:Coalescence_rot_Mom}
\end{figure}

In order to capture the relative motion of the bubbles with respect to each other, it is important to suppress the coalescence of bubbles compared to their relative motion. To illustrate the capability of the present model in this regard, in the last benchmark test, a system of two coalescing bubbles is simulated in the presence of a rotational flow for two different interface energies, $\sigma_{2} = 10^{-2} \sigma_{1}$ and $\sigma_{3} = 10^{-4} \sigma_{1}$. Figure.~\ref{fig:Coalescence_rot_Mom} shows the results of these simulations at three different times. Each row corresponds to the same time starting from top as initial time ($t=0$) to bottom as the final time. For the simulation with the larger interface energy ($\sigma_{2}=10^{-2}\sigma_1$, left column), the two bubbles partially merge as they rotate about $1/8$ of a cycle in an anticlockwise manner, Figure.~\ref{fig:Coalescence_rot_Mom}-a. On the other hand, for the simulation with a 100 times lower interface energy ($\sigma_{3}=10^{-4}\sigma_{1}$, right column) hardly any advance in the coalescence of bubbles is observed for the same amount of rotation, Figure.~\ref{fig:Coalescence_rot_Mom}-b. This clearly demonstrates the maturity of the present model in controlling (slowing down of) the rate of coalescence compared to other concurrent processes. It is noteworthy that in Fig.~\ref{fig:Coalescence_rot_Mom}-b, the attached bubbles deform slightly from their circular shape. This is a result of using a very low interface energy, which makes it difficult for the bubbles to resist the deformation induced by the shear forces of the flow.

\section{Conclusion and outlook}
In this work, we present a strictly mass conserving simple phase-field based model for simulation of liquids containing gas bubbles. The work is motivated by the need to better understand structure formation in metallic foams. A central requirement, therefore, is the capability of the model to slow down the rate of coalescence as compared to bubble rearrangement dynamics. Through a number of carefully selected benchmark tests, the validity of the model is first shown. In a further step, the model is applied to coalescence of two coalescing bubbles demonstrating that different rates of coalescence can indeed be achieved. Most importantly, it is shown that the model allows to study rearrangement of bubbles (induced here via a rotational flow) while suppressing the merging process to a large extent. However, since the rate of coalescence is reduced at the expense of a low interface free energy, bubbles tend to deform correspondingly more easily. As topic for the future work, it would  be desirable to also control bubble formability independent of the merging rate.

\section*{Acknowledgments}
\label{sect:acks}
This work was performed with support from the IMPRS-SurMat programme. The authors would like to acknowledge financial support from ThyssenKrupp AG, Bayer Material Science AG, Salzgitter Mannesmann Forschung GmbH, Robert Bosch GmbH, Benteler Stahl/Rohr GmbH, Bayer Technology Services GmbH, and the state of North Rhine–Westphalia, as well as the European Union in the framework of the ERDF.

\bibliographystyle{elsarticle-num}
\bibliography{mybibfile}

\end{document}